\documentclass[aps,prb,twocolumn,superscriptaddress,showpacs]{revtex4}

\usepackage{amsmath,bm}
\usepackage{graphicx}

\begin{document}

\title{Encapsulation and polymerization of acetylene molecules inside a carbon nanotube}
\author{Gunn Kim}
\affiliation{BK21 Physics Research Division, Seoul National University, Seoul 151-747, Korea}
\author{Yeonju Kim}
\affiliation{School of Physics, Seoul National University, Seoul 151-747, Korea}
\author{Jisoon Ihm}
\email[corresponding author. E-mail:\ ]{jihm@snu.ac.kr}
\affiliation{School of Physics, Seoul National University, Seoul 151-747, Korea}

\begin{abstract}
We study the energetics of acetylene (${\rm C_2H_2}$) molecules 
inside a carbon nanotube (CNT) using the {\it ab initio} pseudopotential method. 
The encapsulation energy of a single ${\rm C_2H_2}$ molecule into the nanotube
and the formation energy of ${\rm (C_2H_2)}_n$@CNT are calculated. 
We investigate whether a polyacetylene chain can be produced 
by fusion of the close-packed acetylene molecules inside the CNT 
and find that there is practically no activation barrier to polymerization.
We propose to employ this method to obtain straight, perfectly isolated, 
and single-stranded polyacetylene chains encapsulated 
inside CNTs, which may be used for molecular electronic devices.

\end{abstract}
\maketitle

\section{Introduction}
Due to the presence of the inner hollow space, the CNT can be an
ideal container for atoms or small molecules. Recently, Takenobu and
coworkers reported that organic molecules predominantly occupy the
inner space of the nanotubes~\cite{Takenobu}. Atoms and molecules
encaged in the nanotube are isolated from the environment and their
chemical reaction could be triggered by electron beam irradiation.
Among alkynes (organic molecules with carbon triple bonds),
acetylene (C$_2$H$_2$) is the simplest and a good model
system to study pressure-induced polymerization which arises at
pressures above 3 GPa at room temperature \cite{Aoki,Santoro}. On
the other hand, under pressure or in the presence of copper, it
can be converted to carbon and hydrogen with explosion.
This instability causes problems in the handling and storage of the
material and the commercial acetylene is stabilized with
acetone and phosphine to prevent explosion. 

McIntosh and coworkers proposed to insert the polyacetylene (PA) 
chains inside carbon
nanotubes and avoid interchain coupling and complicated morphology
of the bulk materials \cite{McIntosh1,McIntosh2}. Because of the
entropic effect, however, the PA (polymer) is not straight and
cannot reach the deep inside of the CNT at room temperature. Here we
propose a method to store acetylene molecules and convert them into
PA using the CNT as a chemical reactor as well as a container.
Through this process, we can obtain a straight, perfectly isolated,
and single-stranded PA encapsulated inside a CNT. Carbon fullerenes
were already found to be converted into a CNT inside the host
nanotube by the coalescence process \cite{Luzzi,Bandow}. A
collection of PA chains under such a morphological control may be
used for future molecular electronic devices. In this work, we
calculate the energetics of acetylene insertion into a nanotube
and find that the reaction is exothermic resulting in a static
capillary force \cite{McIntosh1,Yoon}. This force compresses
incorporated acetylene molecules with an effective pressure of the
order of sub-GPa. Once acetylene molecules are aligned compactly
inside the tube, polymerization of them is predicted to take place
practically without an activation energy barrier.

\section{Computational Details}
We perform {\it ab initio} electronic structure calculations for
C$_2$H$_2$ molecules inside the CNT based on the density functional
theory (DFT) within the Local Density Approximation for the
exchange-correlation energy. We also checked the energy shift by
employing the Generalized Gradient Approximation for some important
geometries, and found that overall results are the same. For the
large supercell calculations, norm-conserving pseudopotentials
\cite{Troullier} are employed in the Kleinman-Bylander
form~\cite{Kleinman}. The wave functions are expanded using the
numerical atomic orbital basis set \cite{openmx1,openmx2} and an
energy cutoff of 100 Ry. 
To calculate the activation energy barrier
in polymerization of the acetylene molecules which require heavier
computational loads, the nudged elastic band (NEB)
method~\cite{neb1,neb2} implemented in the PWSCF
package~\cite{pwscf} is used. Ionic cores are described by ultrasoft
pseudopotentials and the Kohn-Sham orbitals are expanded in plane
waves with a kinetic energy cutoff of 30 Ry.
We have done test calculations for diamond and C$_{60}$,
and obtained converged results for electronic and structural properties with 
two cutoff energies, 100 and 30 Ry, for norm-conserving 
and ultrasoft pseudopotentials, respectively.
The NEB method enables us to find the optimal reaction pathway 
with the minimum energy when both the initial and final states are known. 
Seven replicas are chosen including the initial and final ones 
to construct the elastic band in this method.
The atomic positions of the CNT, acetylene, and polyacetylene chain are relaxed, respectively 
until the atomic force becomes less than 0.05 eV/\AA.

\section{Results and Discussion}
We first discuss the energetics of ${\rm C_2H_2}$ insertion into a
CNT. The dangling bonds of the open (5,5) nanotube of 2 nm in length
are passivated by hydrogen atoms and the ${\rm C_2H_2}$ molecular
axis is assumed to be initially aligned with the tube axis as shown
in Fig. 1(a). The supercell size in the lateral ({\it xy}) direction
is chosen to be as large as 20 \AA~ to avoid interaction between
(5,5) CNTs with $\sim$ 7 \AA~ in diameter 
and that in the axial direction ({\it z}-axis) is 70
\AA~to allow for sufficient vacuum regions. The abscissa in Fig.
1(b) represents the position of the center of C$_2$H$_2$ from the
edge carbon atoms of the CNT along the tube axis. A static capillary
force associated with the energy gain during encapsulation may
compress other ${\rm C_2H_2}$ molecules inside the CNT. The energy
lowering by the axial insertion of an acetylene molecule into the
(5,5) CNT is $\approx$ 0.65 eV as shown in Fig. 1(b), indicating an
exothermic reaction. The energy slope along the tube axis in Fig.
1(b) corresponds to a static capillary force of $\approx$ 0.22 nN.
The energy is further lowered by 0.1 eV when ${\rm C_2H_2}$ is
tilted by about 45$^\circ$ with respect to the axis of the (5,5) nanotube.
In this situation, the distance between a H atom on C$_2$H$_2$ and
the tube wall is $\approx$ 2.3 \AA. 
Because of the limitation of the computational resource, 
the data presented in Fig. 1(b) are obtained without optimization for the whole system.
According to our test calculation, the full relaxation
lowers the total energy by only 0.03 eV (30 meV).  
Thus we believe that we can do without complete optimization
for the supercell containing the whole system 
(i.e., the nanotube and acetylene molecule).

Small but finite energy variation for positive values of the abscissa 
in Fig. 1(b) implies that the acetylene molecule inside the tube 
feels the lattice structure of the CNT.
The periodicity of the variation is almost the same as that of the armchair CNT ($\sim$ 2.5 \AA).
In addition, the tilting of the encapsulated C$_2$H$_2$ molecule indicates the nonbonding interaction
between the nanotube wall and C$_2$H$_2$. 
The recent density functional
calculations show that the long organic molecules such as
tetracyanop-quinodimethane (TCNQ) and
tetrafluorocyanop-quinodimethane (F4TCNQ) encapsulated in the bigger
 CNT are tilted by about 30$^\circ$ with respect to the tube axis \cite{Jing-Lu,Liang}.
\begin{figure}[t]
  \centering
  \includegraphics[angle=0,width=8.5cm]{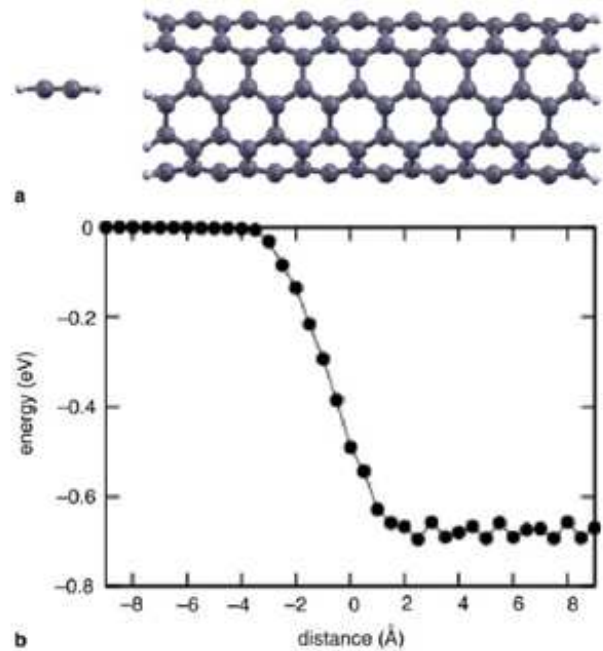}
  \caption{
Energetics of acetylene insertion into a CNT. (a) Atomic model of a
C$_2$H$_2$ molecule entering a (5,5) nanotube. (b) Energy profile
during the insertion of C$_2$H$_2$ along the tube axis.
  }
\end{figure}

We then calculate the activation energy barrier along the reaction
path between two locally stable configurations using the NEB method.
Two stable configurations are the one-dimensional close-packed array
of C$_2$H$_2$ molecules (Geometry A) and the PA chain (Geometry B)
inside the (5,5) CNT, respectively displayed in Fig. 2(a) and (b).
As mentioned above, the acetylene molecule prefers to be tilted by
45$^\circ$ to the tube axis inside the (5,5) CNT (Geometry A). First
of all, the formation energy of Geometry A is calculated. We define
the formation energy of the close-packed array of acetylene
molecules in the (5,5) CNT,
\begin{equation}
E_{form}=E[{\rm CNT+(C}_2{\rm H}_2)_n]-E[{\rm CNT}]-E[({\rm C_2H_2})_n],
\end{equation}
where $E[{\rm CNT}]$ and $E[({\rm C_2H_2})_n]$ are the energies of the
isolated nanotube and C$_2$H$_2$ while $E[{\rm CNT+(C}_2{\rm H}_2)_n]$ is the
energy of the whole encapsulated system.
For the plane-wave calculation, the unit cell size in the lateral ({\it xy}) direction
is chosen to be 15 \AA~ and that in the axial direction ({\it z}-axis) is 2.5 \AA~
with the periodic boundary condition.
The formation energy is $-$0.4 eV per C$_2$H$_2$ (exothermic reaction). 
We do not consider a Peierls distortion in calculating
the energy of the PA chain. Figure 2(c) reveals that Geometry B has
a lower energy than Geometry A by 4.3 eV per C$_2$H$_2$ and there is
no activation barrier. In this reaction process, a $\pi$-electron in
the triple bonding rapidly hops to form a new carbon-carbon bond
with the nearest-neighbor C$_2$H$_2$. The initial and final dots in
(c) indicate the energies of Geometries A and B, respectively.
Therefore, close-packed acetylene molecules are driven to be
polymerized by fusion in the tube. During the process of
polymerization, the carbon-carbon bonds of the acetylene molecule
are elongated (from 1.2 to 1.4 \AA) and the angles between carbon
and hydrogen atoms are changed.
\begin{figure}[t]
  \centering
  \includegraphics[angle=0,width=8.5cm]{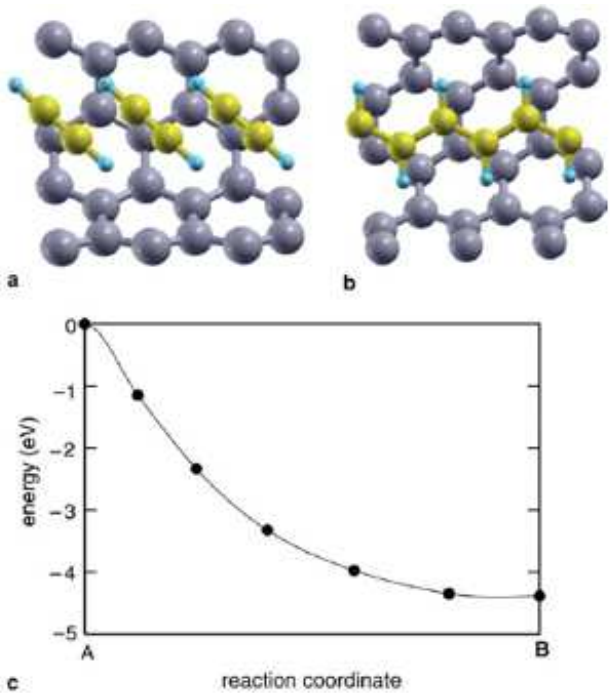}
  \caption{
(color online) Polymerization process of acetylene in the CNT. (a) Atomic model of
a one-dimensional close-packed array of acetylene molecules inside a
(5,5) CNT (Geometry A). (b) Atomic model of the {\it
trans}-polyacetylene formed inside a (5,5) CNT (Geometry B). (c)
Energy diagram along the reaction pathway of polymerization from
Geometry A to B.
  }
\end{figure}
We have also tested the acetylene encapsulation and polymerization
inside a larger diameter nanotube, (6,6) CNT. Insertion of
C$_2$H$_2$ into the (6,6) tube is more favorable than that of the
(5,5) tube energetically. However, the (6,6) tube has a region of a
constant potential inside \cite{McIntosh1} and the randomness caused
by the lateral freedom of the C$_2$H$_2$ position in that region
(owing to the entropic effect) makes polymerization a little bit less
favorable. For dense C$_2$H$_2$ molecules in the (6,6) tube, the
equilibrium angle of the C$_2$H$_2$ axis with the tube axis is found
to be 90$^\circ$ before polymerization. Other than these
differences, we obtain similar energetics to the (5,5) tube case,
including the absence of the activation barrier to polymerization.

Even if there is practically no activation energy barrier, it is
very hard for the molecules to react with the neighbor molecule spontaneously inside the CNT
before arranging themselves into a one-dimensional close-packed array (Geometry A).
The isomers of C$_4$H$_4$ cannot be created from acetylene molecules
at room temperature because C$_2$H$_2$ molecules are not
dissociated. In addition, some isomers of C$_4$H$_4$ such as
cyclobutadien (ring-shaped) are energetically unstable. When the
encapsulated C$_2$H$_2$ molecules form the one-dimensional
close-packed array by the effective pressure of the order of sub-GPa
associated with the capillary force, polymerization of acetylene
takes place without activation energy barrier.
Interestingly, it is known that the equilibrium spacing between C$_{60}$ molecules in the 
nanotube is smaller by $\sim$ 4 \% than in the three-dimensional C$_{60}$ crystal 
by the high resolution transmission electron microscopy (HRTEM) \cite{Smith}, 
electron diffraction \cite{Hirahara1}, and Raman measurements \cite{Hirahara2}.
It indicates that incorporated molecules inside the CNT may be subject to an axial compressive force
due to the energy change along the tube axis during the insertion \cite{Yoon}.
In the same manner, acetylene molecules may be close-packed in the nanotube.

\section{Conclusion}
We have performed {\it ab initio} pseudopotential calculations to
investigate the energetics of acetylene (${\rm C_2H_2}$) molecules
inside a CNT. The encapsulation process of a single ${\rm C_2H_2}$
molecule in the nanotube is an exothermic reaction and the static
capillary force produced by entering C$_2$H$_2$ is a fraction of
a nanonewton. It is found that there is practically no activation
barrier to the polymerization reaction, implying that a PA chain can
be easily produced inside the CNT by fusion of the close-packed
acetylene molecules in it. Thus, we propose a method to fabricate a
straight and isolated PA chain inside a CNT, by which we can avoid
the interchain coupling and understand the intrinsic properties of
the individual PA chains. In addition, it is a promising way to
investigate and control the interaction between the CNT and the
polymer. We believe that this study sheds light on physical and
chemical properties of organic molecules and leads to an application
of the nanotube as a nano-sized chemical reactor.

\section*{Acknowledgement}
This work was supported by the CNNC of Sungkyunkwan University, the BK21 project of KRF
and the MOST through the NSTP (grant No. M1-0213-04-0001).
The computations were performed at Supercomputing Center of KISTI
through the Supercomputing Application Support Program.

\end{document}